\documentclass[10pt,conference]{IEEEtran}

\usepackage{setspace}
\usepackage{hyperref}
\usepackage{amsmath}
\usepackage[numbers, sort&compress]{natbib}

\usepackage{graphicx}
\usepackage{comment}

\newcommand{\be}{\begin{eqnarray}}
\newcommand{\ee}{\end{eqnarray}}

\usepackage{listings}
\usepackage[font=footnotesize]{caption}
\usepackage[font=footnotesize]{subcaption}
\usepackage{multirow} 

\usepackage{url}

\usepackage{enumitem}

\usepackage{xcolor,soul}

\usepackage[ruled,linesnumbered]{algorithm3e}
\renewcommand{\algorithmautorefname}{Algorithm}
\usepackage{balance}

\newcommand\couldremove[1]{{\st{#1}}}
\newcommand{\remove}[1]{}

\definecolor{lightgray}{gray}{0.6}
\definecolor{lightblue}{rgb}{0.9,0.9,1}
\definecolor{aqua}{rgb}{0.0, 1.0, 1.0}

\newcommand{\hlc}[2][aqua]{{\sethlcolor{#1}\hl{#2}}}
\newcommand\note[1]{\hlc[aqua]{#1}}
\newcommand\lin[1]{\hlc[yellow]{LZ: #1}} 
\newcommand\lina[1]{\hlc[gray]{LZ: #1}}
\newcommand\yue[1]{{\hlc[orange]{YW: #1}}} 
\newcommand\yuea[1]{\hlc[gray]{YW: #1}}
\newcommand\nami[1]{{\hlc[pink]{NL: #1}}}
\newcommand\namia[1]{{\hlc[gray]{NL: #1}}}
\newcommand\emmet[1]{{\hlc[green]{EH:#1}}}

\newcommand\name[0]{{\sc DecoNet}\xspace}
\newcommand\hname[0]{{\sc DecoNet/Helios}\xspace}
\newcommand\block[0]{decoding block\xspace}
\newcommand\blocks[0]{decoding blocks\xspace}
\newcommand\Block[0]{Decoding Block\xspace}
\newcommand\Blocks[0]{Decoding Blocks\xspace}

\newcommand\unit[0]{decoder instance\xspace}
\newcommand\units[0]{decoder instances\xspace}

\newcommand\todo[1]{\textcolor{red}{TODO: #1}}

\renewcommand\couldremove[1]{}  
\renewcommand\note[1]{}         
\renewcommand\lin[1]{}          
\renewcommand\lina[1]{}         
\renewcommand\nami[1]{}          
\renewcommand\namia[1]{}         
\renewcommand\yue[1]{}          
\renewcommand\yuea[1]{}         
\renewcommand\todo[1]{}         
\renewcommand\emmet[1]{}        
\renewcommand{\hlc}[2][aqua]{}

\usepackage{etoolbox}

\begin{document}
%
\title{Network-Integrated Decoding System for Real-Time Quantum Error Correction with Lattice Surgery}


\author{\IEEEauthorblockN{Namitha Liyanage,
Yue Wu, Emmet Houghton, and
Lin Zhong}
\IEEEauthorblockA{Department of Computer Science,
Yale University, 
New Haven, CT\\
Email : \{namitha.liyanage, yue.wu, emmet.houghton, lin.zhong\}@yale.edu}}


%



\maketitle
\thispagestyle{plain}
\pagestyle{plain}

\begin{abstract}
Existing real-time decoders for surface codes are limited to isolated logical qubits and do not support logical operations involving multiple logical qubits. 
We present \name, a first-of-its-kind decoding system that scales to thousands of logical qubits and supports logical operations implemented through lattice surgery. 
\name organizes compute resources in a network-integrated hybrid tree-grid structure, which results in minimal latency increase and no throughput degradation as the system grows. 
Specifically, \name can be scaled to any arbitrary number of $l$ logical qubits by increasing the compute resources by $O(l\times log(l))$, which provides the required
$O(l)$ growth in I/O resources while incurring only an $O(log(l))$ increase in latency—a modest growth that is sufficient for thousands of logical qubits. 
Moreover, we analytically show that the scaling approach preserves throughput, keeping \name backlog-free for any number of logical qubits
We report an exploratory prototype of \name, called \hname, built with five VMK-180 FPGAs, that successfully decodes 100 logical qubits of distance five. For 100 logical qubits, under a phenomenological noise rate of 0.1\%, the \hname has an average latency of 2.40 $\mu$s and an inverse throughput of 0.84 $\mu$s per measurement round. 
\end{abstract}


%
\IEEEpeerreviewmaketitle

\section{Introduction}

Quantum error correction (QEC) is essential for fault-tolerant quantum computing (FTQC). 
QEC employs multiple physical qubits to form a single, more resilient logical qubit, such as the surface code~\cite{dennis2002topological, fowler2012surface}. 
A decoder identifies potential errors with these physical qubits before they lead to a logical error.
\emph{Lattice Surgery}~\cite{horsman2012surface} is a popular approach for implementing logical operations with logical qubits in surface codes. It merges and splits logical qubits, avoiding the need for non-local interactions but creating fresh challenges for the decoder design. See \S\ref{sec:background}.
Although significant progress has been made in decoder development~\cite{battistel2023realtime, campbell2024series}, all existing real-time decoder implementations are limited by the resources available from a single node (FPGA or computer). As a result, they support isolated logical qubits~\cite{das2022afs, higgott2023sparse, Vittal2023Astrea, barber2025realtime, Liao2023WitGreedy,das2022lilliput, liyanage2024fpga, wu2024micro}. 
To the best of our knowledge, no real-time decoder implementation supports QEC with lattice surgery. See \S\ref{sec:related}.

In this paper, we report \textbf{\name}, a scalable real-time quantum error correction system that overcomes resource limitations by utilizing multiple network-integrated FPGA nodes.
\name distributes the decoding workload across several network-integrated compute resources, supporting larger code distances and multiple logical qubits.
\name assigns a decoder instance for each logical qubit, which first decodes the logical qubit in isolation. The system then combines the results of the decoder instances to solve the global decoding problem. 

\name organizes compute resources in a network of a hybrid tree-grid topology, allowing both computational power and I/O capacity to scale by adding FPGA nodes over the network. As a result,
\name can decode large-scale circuits containing hundreds of logical qubits interacting through lattice surgery operations. 
Furthermore, \name can decode circuits with dynamic interactions, such as those involving conditional operations. 
No prior real-time, graph-based decoder can handle dynamic configurations, as they assume a static decoding graph (See \S\ref{ss:decoders}).
However, practical quantum algorithms involve conditional gates, resulting in a dynamic decoding graph that must be generated at runtime. \name efficiently decodes these dynamic graphs using a fusion operation.

We have prototyped \name using the Helios decoder~\cite{liyanage2024fpga} as \units. The system, \hname, consists of five Xilinx VMK-180 FPGA evaluation boards~\cite{vmk180} arranged in a two-level tree topology. 
Additionally, \hname features a custom low-latency interconnect for the four leaf nodes.
We experimentally show that \hname can decode up to 100 logical qubits at code distance 5, including realistic lattice surgery operations. 
For 100 logical qubits, \hname achieves an average decoding latency of 2.40~$\mu s$ and an inverse throughput of 839~ns per measurement round under 0.1\% phenomenological noise.
This result demonstrates, for the first time, the feasibility of parallel decoding for 100 interconnected logical qubits in real time.

In summary, we report the following contributions.

\begin{itemize} [leftmargin=*]
\item The first scalable real-time QEC decoder capable of decoding thousands of logical qubits by distributing workloads across multiple network-integrated FPGA nodes, breaking the single-node resource constraint of previous decoders. 
\item The first real-time decoder capable of decoding lattice-surgery-based dynamic decoding graphs 
\item A set of empirical data demonstrating that \name, implemented using five FPGAs, can successfully decode 100 logical qubits with lattice surgery operations in real-time, a first for a decoder implementation. 
\item Fusion Union-Find, a novel method for parallelizing the decoding of large, dynamic decoding graphs, which is 25-40\% faster than parallel window decoding.
\end{itemize}

\section{Background}
\label{sec:background}

We provide brief overviews of the implementation of logical operations using lattice surgery, the requirements for decoders to support such operations, and the role of qubit controllers.

\subsection{Lattice Surgery}

\begin{figure} [!t]
\centering
\includegraphics[width=0.99\linewidth]{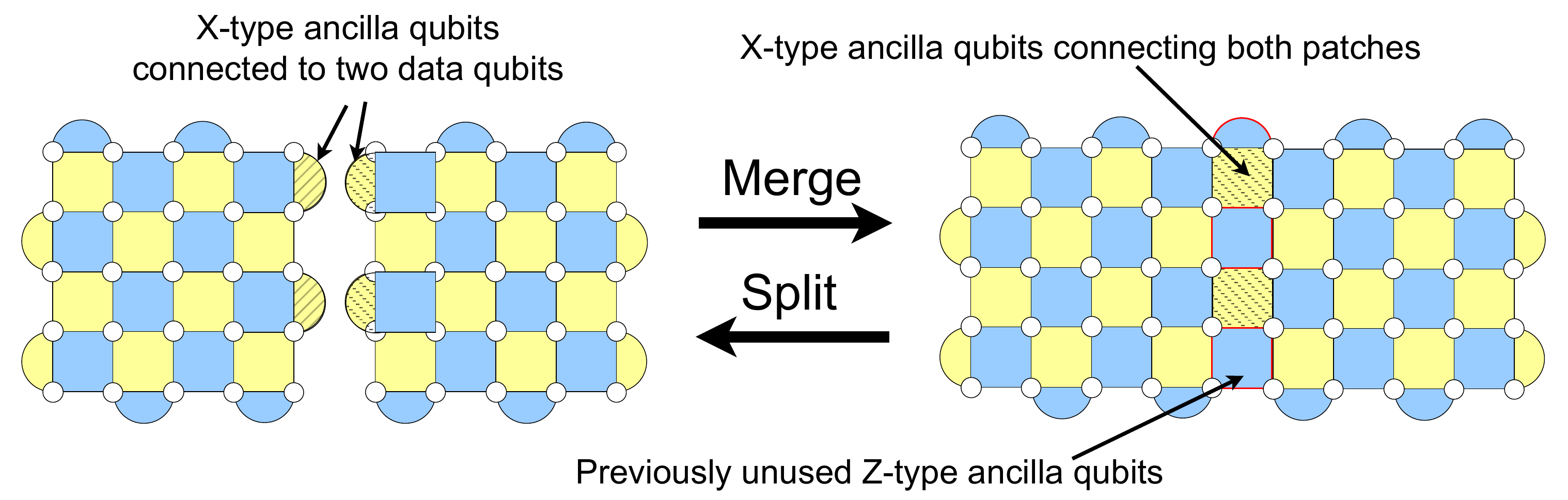}
\caption{Merging of two $d=5$ surface code patches along their $Z_L$ boundaries involves measuring joint ancillas across the boundary to form a single logical patch.
Specifically, X-type ancilla qubits connected to two data qubits at the boundary (marked by arrows in the top left) are extended into ancilla qubits interacting with data qubits from both patches (arrows in the top right). Merging also activates previously unused Z-type ancilla qubits at the boundary, whose joint measurement yields the $Z_L Z_L$ operator. The newly activated Z-type ancillas in the merged qubit are shown with red borders.}
\label{fig:latticesurgery}
\end{figure}

Lattice surgery~\cite{horsman2012surface} is a resource-efficient method to implement logical operations with more than one logical qubits encoded in surface codes.
As lattice surgery requires only nearest-neighbor physical qubit interactions, it enables logical circuits to be executed on planar quantum hardware without the need for long-range connections between qubits. 
This hardware-friendly property makes lattice surgery a promising candidate for scalable fault-tolerant quantum computing. 
As a result, researchers have embraced lattice surgery over alternative methods such as transversal gates~\cite{Lao2019mapping, Watkins2024highperformance, lin2024spatially, bombin2023modular, ueno2022qulatis}.

Lattice surgery involves \emph{merging} and \emph{splitting} surface code patches to perform multi-qubit logical operations. 
A merge operation joins the adjacent boundaries of two surface code patches by measuring ancilla qubits that interact with data qubits from both logical qubits across the boundary, as illustrated in \autoref{fig:latticesurgery}.
Depending on whether the $X_L$ or $Z_L$ boundary is merged, this process consists of two concurrent steps: (1) applying additional gates between data qubits at the boundary of one logical qubit and ancilla qubits at the boundary of the other, (2) measuring previously unused ancilla qubits at the boundary. 
In \autoref{fig:latticesurgery}, merging occurs along the $Z_L$ boundary: it converts the X-type ancilla qubits at the boundary into ancilla qubits connecting both logical qubits, and measures previously unused Z-type ancillas (highlighted with red borders). 
This approach is generalizable to more than two logical qubits, and two distant logical qubits can be merged using a chain of ancillary logical qubits~\cite{horsman2012surface}.

Splitting is the reverse operation of merging. After merging, the combined logical patch can be split back into two separate logical qubits by ceasing joint stabilizer measurements and resuming individual stabilizer measurements for each patch.

Lattice surgery operations, especially merge, pose additional challenges to QEC decoding. 
The merging of logical qubits results in shared error syndromes between the qubits. 
Since any two logical qubits in the system can potentially be merged, the worst-case scenario requires the decoder to handle a merged patch that spans the entire qubit array.
This necessitates a decoder capable of processing all logical qubits collectively, as independent decoding of each logical qubit would fail to accurately decode the shared error syndromes introduced by merging. 

In comparison, splitting is much simpler. Once a patch is split, two decoders can decode the resulting logical qubits independently. 
Decoders can exploit this parallelism to improve decoder throughput when managing multiple logical qubits.

\subsection{QEC Decoders}
\label{ss:decoders}

\begin{figure} [!t]
	\begin{subfigure}{0.32\linewidth}
        \includegraphics[width=0.99\textwidth]{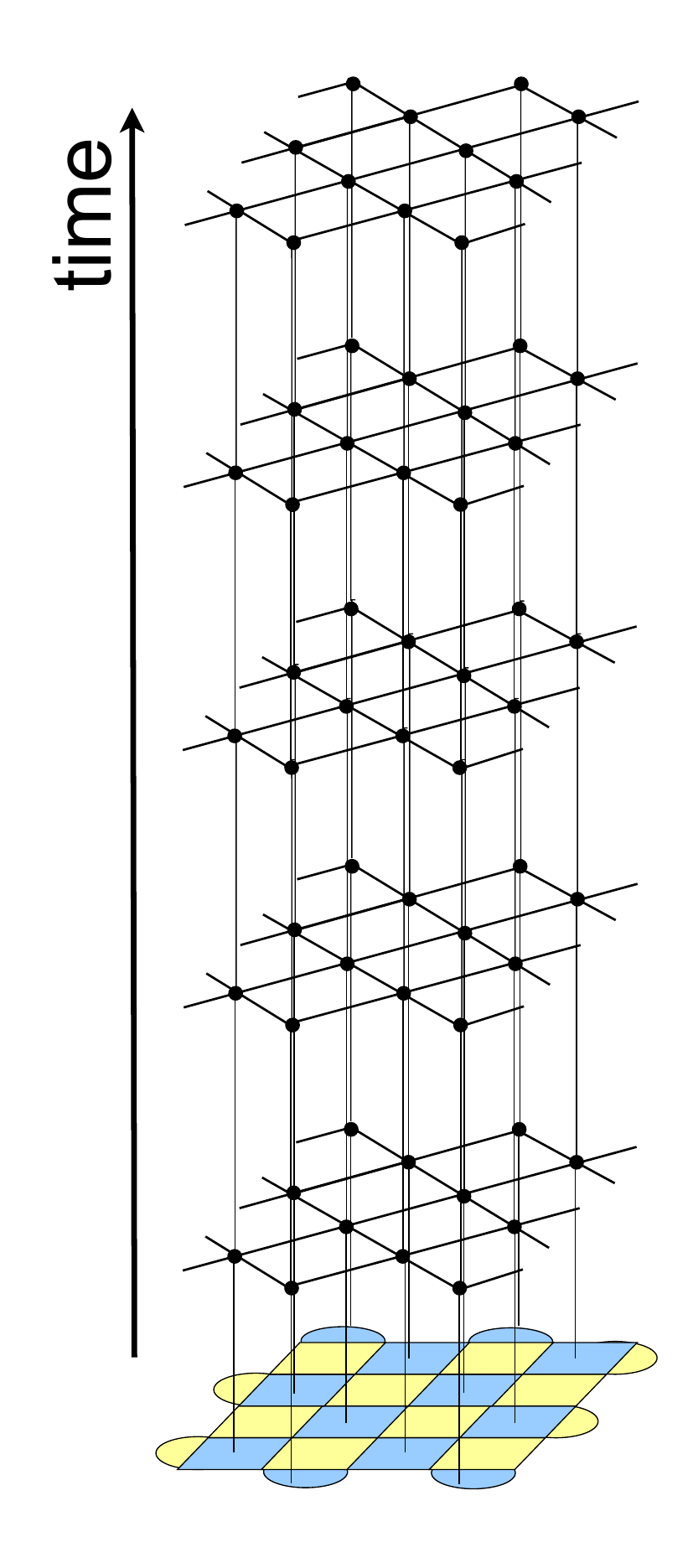}
        \label{fig:one_qubit_decoding}
    \end{subfigure}    
	\begin{subfigure}{0.60\linewidth}
        \includegraphics[width=0.99\textwidth]{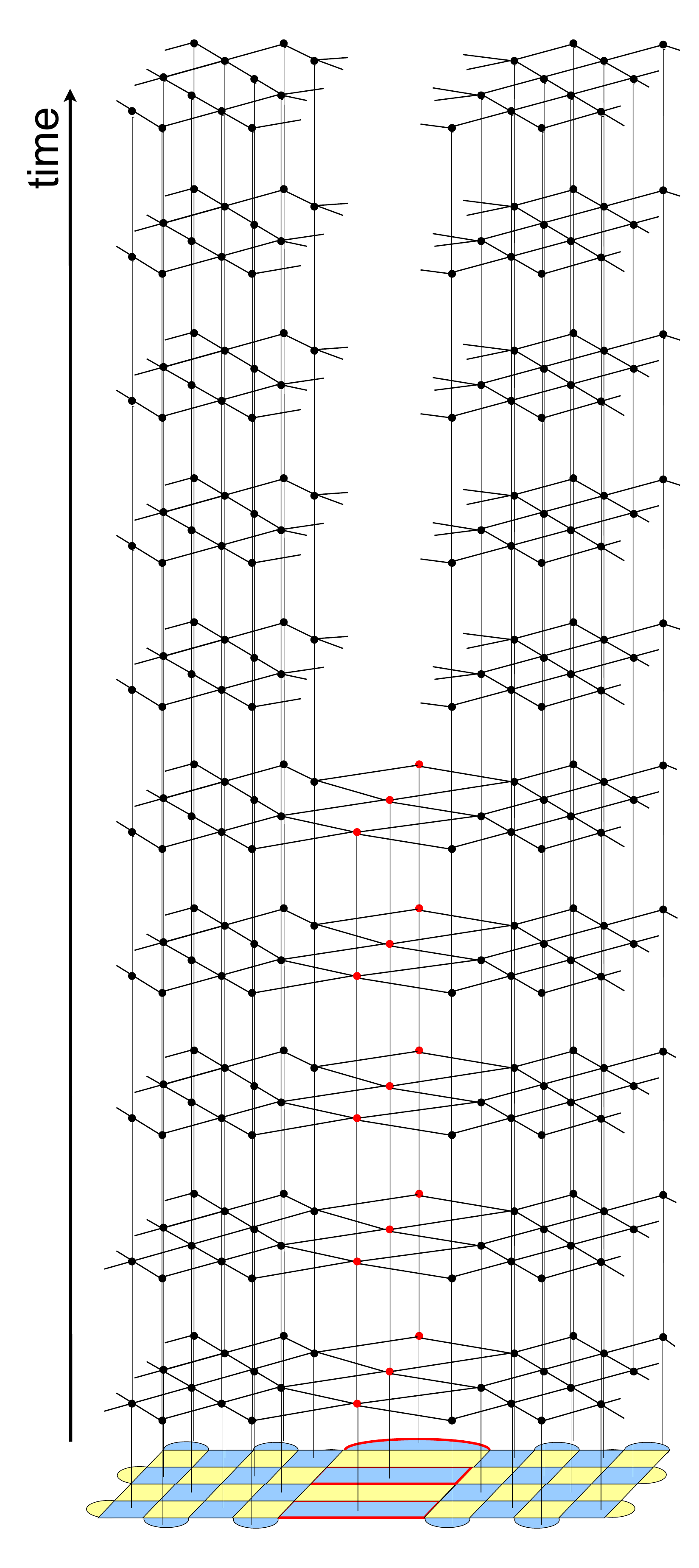}
        \label{fig:two_qubit_decoding}
    \end{subfigure}
	\caption{Lattice surgery modifies the structure of the decoding graph, assuming under a phenomenological noise model for a $d=5$ surface code. (top) Decoding graph of Z-ancillas over five measurement rounds. (bottom) Decoding graph of a system involving two logical qubits that are merged for five rounds and then split. The decoding graph contains additional vertices (marked in red) when the two logical qubits are merged.
    }
	\label{fig:decode_graph}
\end{figure}

When measurements of the ancilla qubits are ready, a decoder, using classical computing, identifies the most likely set of physical errors that could cause the observed defect measurements.
To mitigate the effects of measurement errors, the decoder processes at least $d$ rounds of measurements together, collectively referred to as a syndrome. 
With lattice surgery, the decoder must decode $d$ rounds of measurements together after each merge or split operation~\cite{horsman2012surface}.

Given the syndrome, the problem of finding the most likely errors is commonly represented by a \emph{decoding graph}. Each vertex in the decoding graph corresponds to an ancilla measurement, and each edge represents a potential error.  
\autoref{fig:decode_graph} (top) illustrates the decoding graph for a single logical qubit coded in a surface code with $d=5$.
For example, Sparse Blossom~\cite{higgott2023sparse} and Parity Blossom~\cite{wu2023fusion} solve this problem exactly, while Union-Find (UF)~\cite{delfosse2021almost} solves this problem approximately~\cite{wu2022interpretation} but faster.

\subsubsection{Dynamic Nature of Decoding Graphs}

With lattice surgery, the decoding graph becomes \emph{dynamic}, because the graph structure depends on the outcomes of previous measurements and conditional operations in the \couldremove{logical }circuit. 
For instance, whether two logical qubits should be merged may be determined by the result of a measurement on a third qubit, only available at runtime, as in the case of a T-gate implemented using magic state injection~\cite{Litinski2019game}. 
As a result, the complete decoding graph cannot be precomputed, and the decoder must construct the decoding graph as the computation progresses. 

This runtime graph construction poses new challenges to decoding. In particular, the decoder must not only process syndrome data in real time, but also update the decoding graph itself based on conditional control flow. This requires tight integration with the control stack and minimal latency to keep up with real-time circuit execution. In \autoref{fig:decode_graph} (bottom), two logical qubits are merged for $d$ rounds and then split apart. Whether this merge occurs can depend on a prior measurement result, meaning that the decoding graph, especially the inclusion of vertices corresponding to shared ancillas during the merge stage (shown in red), is not known ahead of time.

\subsubsection{Performance Metrics}

The performance of a decoder is evaluated using three metrics: accuracy, throughput, and latency. \emph{Accuracy} determines the size of the surface codes to achieve a given logical error rate. 
Decoders with lower accuracy require much more hardware. 
For current physical error rates in superconducting quantum computers, $p=0.001$, a Union-Find decoder requires $d \approx 29$ to achieve a logical error rate of $10^{-15}$.
\emph{Latency} determines the rate of logical operations, as intermediate decoding results are often needed for time-sensitive operations such as T-gates~\cite{Terhal2015quantum}. 
As a result, latency determines the overall execution time of the circuit and the probability of failure. 
For superconducting quantum computers, researchers typically assume a decoding latency budget of $10\ \mu s$ when estimating the timing of quantum algorithms~\cite{Gidney2021rsa}.
\emph{Throughput} or rate of decoding must match the measurement rate. Otherwise, a backlog of undecoded measurements can accumulate, exponentially slowing down the quantum computer. For superconducting quantum computing, which has the most stringent requirements, the decoding rate should be $1 \mu s$ per measurement round~\cite{Chen2021Exponential}.

In this context, a \emph{real-time} decoder is one whose throughput exceeds the measurement rate and whose latency is comparable to the time to measure $d$ rounds. 
For state-of-the-art superconducting quantum systems, this corresponds to a throughput greater than one million measurements per second and a total latency of approximately $d\ \mu s$~\cite{zaid2024local}.



\subsection{Qubit Controller}
In a quantum computer, the decoder must receive measurements from the qubit controllers.
Using either an FPGA or specialized hardware, a controller generates the control signal to manipulate a qubit. A quantum computer needs 100s of qubits and therefore 100s of controllers in parallel to support experiments with many logical qubits~\cite{ofek2016extending,Xu2023QubiC2,ryan2016hardware,google2019quantum,QuantumMachines,pqsc2025,qblox2023}.

This poses specific design requirements for the decoder. First, the decoder must be tightly integrated with qubit controllers to aggregate data from all the controllers with minimal latency. 
Second, the decoder's compute and I/O communication capabilities must scale with the number of qubit controllers to prevent it from becoming a system bottleneck.

\section{Related work}
\label{sec:related}

To our knowledge, \hname is the first scalable real-time decoder system that provides empirical results on decoding hundreds of interacting logical qubits simultaneously, including support for dynamic lattice surgery circuits.
\hname draws inspiration from several works that explored decoding lattice surgery-based dynamic graphs.
It is the first decoder system that is capable of using network-integrated compute resources and supports real-time decoding with thousands of logical qubits. 

Wu et al.~\cite{wu2024lego} suggest a distributed decoding system in which a coordinator assigns decoding blocks to multiple compute resources and merges them using a fusion operation. \hname can be considered the first implementation of this distributed decoding system, with new implementation ideas such as combining fusion with windowing to reduce latency and statically assigning decoding blocks to hardware units for more efficient execution.
Bombin et al.~\cite{bombin2023modular} propose a modular decoding framework that partitions the decoding graph into subgraphs. These subgraphs correspond to commonly used logical operations. 
The framework first decodes the boundaries between these subgraphs, followed by parallel decoding of the individual subgraphs. 
The authors also considered dynamic circuits and proposed a hardware architecture in which multiple processing units, sharing a global memory, decode subgraphs in parallel, without providing a hardware implementation or any empirical data on decoding latency or throughput.

Lin et al.~\cite{lin2024spatially}, Skoric et al.~\cite{skoric2023parallel}, and Tan et al.~\cite{tan2022scalable} propose partitioning the decoding graph into spatial regions, or windows, that can be decoded in parallel with limited inter-window dependencies. 
Lin et al. further analytically show that inter-window communication latency does not impose a scalability bottleneck. 
Our work builds on both these efforts to implement window-based decoding across FPGAs and, for the first time, \emph{empirically} validating that inter-FPGA communication is not a scalability bottleneck. 

To our knowledge, QULATIS~\cite{ueno2022qulatis} is the only reported hardware \emph{design} that supports lattice surgery-based dynamic decoding graphs. 
It decodes errors using a greedy algorithm that sacrifices accuracy for simplicity. 
Without an implementation, the authors evaluated the decoding latency of QUALTIS using SPICE simulations.
Compared to QUALTIS, \hname features a much more scalable architecture that leverages network-integrated FPGAs. Furthermore, our implementation of \hname employs a Union-Find-based decoder, achieving at least two orders of magnitude higher decoding accuracy compared to the greedy algorithm used in QULATIS.

\section{\name Overview}
\label{sec:Design}

\subsection{Design Goals}

Our goal is to build a system capable of real-time decoding of multiple interacting logical qubits, implemented by the surface code. 
In graph-based decoding, this means handling a dynamic decoding graph comprising all logical qubits in the quantum computer. 
The system should be both scalable and adaptable. 
\emph{Scalability} refers to the system's capability to handle increasingly larger decoding graphs by incrementally and trivially adding compute resources without significant redesign. 
\emph{Adaptability} refers to the system's ability to support a dynamic decoding graph whose complete structure is not fully known before runtime. 

\subsection{Key Ideas}

\name achieves its design goals by integrating several key ideas:

 \emph{Partitioning the decoding graph into \blocks:} 
Since a monolithic decoder cannot scale efficiently due to hardware resource constraints, we partition the decoding graph into smaller units independently decodable, called decoding blocks, as inspired by~\cite{wu2024lego}. 
The system partially decodes each decoding block separately using a decoder and combines them to form a global decoding solution.

\emph{Network-Integrated Resources for Scalability:}
\name uses network-integrated compute resources to scale beyond the limitations of a single node, computer, or FPGA.
Unlike board- or chip-level integration, network integration allows us to incrementally and trivially add resources to scale the design.

 \emph{Fusion-based and Window-based Approaches:} We use two complementary methods to combine decoding blocks into a global decoding solution. Within a single computational resource (such as a single FPGA), we use a fusion-based approach~\cite{wu2023fusion, wu2024lego}.  Across multiple network-integrated compute resources, we adopt a window-based approach~\cite{lin2024spatially}.  We detail these methods and our partitioning strategy in \S\ref{sec:parallelsim}.

\emph{Hybrid Tree-Grid Network Topology:} To minimize communication latency among compute resources, we organize the compute resources in a novel hybrid topology which we describe in detail in \S\ref{sec:hardware}. The hybrid topology uses a grid structure to facilitate local communication required for lattice surgery operations among neighboring compute resources, and a tree structure to ensure minimal worst-case latency when communicating decoding outcomes between compute resources. Additionally, the hybrid topology allows I/O resources to scale proportionally with the number of compute resources.

The design of \name is agnostic to the choice of decoder implementation. This flexibility is crucial as decoding technologies rapidly evolve, each offering different trade-offs in accuracy, latency, and resource utilization. \name only poses two requirements to the decoder: first, it must decode a \block of $\approx d^3$ vertices in real-time; and second, it must support the fusion of \blocks across both space and time.
These requirements are reasonable in view of recent advancements in decoder designs. Many decoders meet the real-time requirement for moderate values of $d$~\cite{das2022afs, higgott2023sparse, Vittal2023Astrea, barber2025realtime, Liao2023WitGreedy, das2022lilliput, liyanage2024fpga, wu2024micro, wu2023fusion, zaid2024local}. Several recent decoders also support fusion~\cite{wu2024micro, wu2023fusion}; both UF and MWPM decoders can be extended to support fusion. 

We implement \name using Helios as \units, referred to as \hname (detailed in \S\ref{sec:implementation}). 
This implementation spans five FPGAs and can decode up to 100 logical qubits at $d=5$.

\section{\name Decoding Algorithm}
\label{sec:parallelsim}

This section describes the decoding algorithm used in \name. 
It outlines how \name partitions the decoding graph into \blocks, distributes them across compute resources to decode them in parallel, and combines them to form the complete graph.

\subsection{\Block}

A \block is the basic unit of decoding in \name. 
For a logical qubit of code distance $d$, a \block consists of $d$ rounds of ancilla measurement outcomes. 
As FTQC requires decoding at least $d$ rounds of measurements after a logical operation, a \block is the smallest subgraph in the decoding graph capable of producing a meaningful result. 

Dividing the decoding graph into \blocks enables fast construction of a dynamic decoding graph.  
Each \block has a static graph structure that \name generates offline.  
At runtime, \name constructs the dynamic decoding graph from these pre-constructed blocks.  
Thus \name represents graph changes such as merges and splits between logical qubits as merges and splits between \blocks.   
By updating only the connections between affected \blocks, \name avoids reconstructing the entire decoding graph at runtime, which would be time-intensive.

Decoding blocks also increases parallelism.  
When logical qubits are split, \name decodes their \blocks in parallel.  
\name combines relevant blocks only when logical qubits merge, after initially decoding them in parallel.

\subsection{Combining \Blocks}
\label{ssec:combining}
We consider two scenarios when combining \blocks: (1) combining within a single compute resource and (2) combining across network-integrated compute resources.


\subsubsection{Intra-resource \Block Fusion}

\name uses a fusion operation previously suggested by Wu et.al.~\cite{wu2023fusion, wu2024lego} to merge blocks within the same compute resource both spatially and temporally.
Fusion provides few advantages compared to alternate combining methods such as parallel window decoding and sliding window decoding.

\begin{itemize} [leftmargin=*]

\item \emph{Reduced Redundant Computation:} Conventional methods require overlapping windows of at least $d/2$ width along every direction to combine blocks, causing redundant computations in the overlapping region. Furthermore, the redundant computation region scales $O(d^3)$.
In contrast, when using a clustering-based decoding algorithm, fusion avoids redundant computations as it preserves the cluster details from the first decoding stage when starting the fusion operation. 
Furthermore, fusion only affects clusters incident to the boundary between merged blocks, which scales $O(d^2)$.

\item \emph{Increased Parallelism:} Conventional methods impose sequential dependencies due to artificial defects along boundaries~\cite{dennis2002topological, lin2024spatially, skoric2023parallel}. 
Fusion eliminates this bottleneck by allowing simultaneous decoding of all the blocks prior to combining. 
This is especially advantageous when merging blocks along the time domain.
When using fusion decoding, the first block can start after the first $d$ rounds are available, whereas sliding window decoders must wait for all $2d$ rounds to start decoding. 

\end{itemize}

\subsubsection{Inter-resource \Block Combination via Parallel Window Decoding}

We use the parallel window decoding method, proposed by Lin et al.~\cite{lin2024spatially} and Skoric et al.~\cite{skoric2023parallel}, to combine \blocks across compute resources due to its lower communication overhead.  
This method reduces communication latency by requiring only one unidirectional message per pair of adjacent decoding blocks, unlike fusion-based methods that involve multiple data exchanges.  
This reduction is critical when combining \blocks across the network, where communication latency often exceeds decoding latency.  
For example, state-of-the-art FPGA decoders can decode a $d{=}21$ surface code in 240~ns~\cite{liyanage2023scalable}, but communication between FPGAs over gigabit transceivers can take around 300~ns~\cite{xilinxaurora}.

The parallel window decoding method partitions the decoding graph into multiple windows~\cite{skoric2023parallel, lin2024spatially}.
In the context of \name, each window consists of \blocks mapped to the same compute resource. 
Parallel window decoding groups these windows into multiple groups such that no two adjacent windows belong to the same group.  
This method then decodes the groups sequentially, performing parallel decoding of windows within each group and transmitting boundary information between groups in order.  
This boundary information represents the most probable physical qubit errors that cross window boundaries.

Prior work shows that at least three groups are required to satisfy the adjacency constraint~\cite{skoric2023parallel, lin2024spatially}.  
Because the number of groups directly impacts decoding latency, \name groups the network-integrated compute resources into three groups to minimize latency while satisfying the constraint.

\subsection{Pipelined Decoding Procedure}

\begin{figure}[!t]
    \centering
    \includegraphics[width=0.99\linewidth]{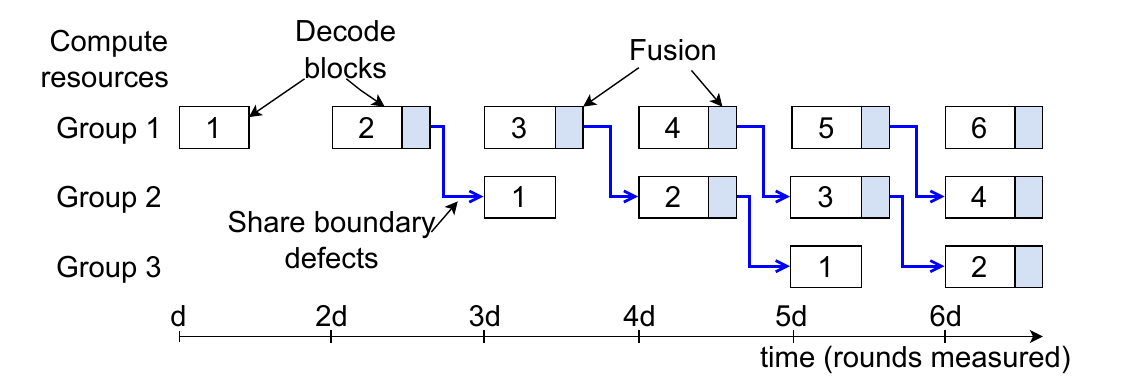}
    \caption{Timeline of the decoding procedure pipelined across three compute resource groups. Each group performs parallel decoding of its assigned \blocks (white boxes). After decoding, \name fuses adjacent \blocks in space and time (blue). The system then shares the boundary information between groups (blue arrows). Numbers in the white boxes denote the round index of \blocks. Groups 2 and 3 lag behind group 1 due to data dependencies.}
    \label{fig:decoding_pipeline}
\end{figure}

Next, we describe the general decoding procedure of \name that organizes the compute resources into three groups as motivated above.  
\autoref{fig:decoding_pipeline} illustrates the pipelined execution of this procedure across the three groups.

When the $d\cdot i^{th}$ round of measurements becomes available, Group~1 decodes the corresponding \blocks assuming they are isolated. After local decoding, Group~1 fuses adjacent \blocks corresponding to merged logical qubits and performs time-domain fusion with \blocks from the $(i-1)^{th}$ round corresponding to the same logical qubit. Once fusion completes, Group~1 commits the $(i-1)^{th}$ round and transmits boundary information to Group~2.
Group~2 then decodes and fuses the $(i-1)^{th}$ round using this information and commits the $(i-2)^{th}$ round. Similarly, Group~3 processes and commits the $(i-3)^{th}$ round.

This pipelined execution introduces a minimum decoding latency of $3d$ rounds but does not affect the throughput of decoding. 
Each group begins decoding a new round of \blocks immediately after it commits the previous round and receives the required boundary information.

\section{\name network architecture}
\label{sec:hardware}

The key to \name's scalability is to exploit network-integrated compute resources, allowing it to go beyond a single compute node, e.g., FPGA, which has limited prior decoder implementations such as Helios~\cite{liyanage2023scalable}, Lilliput~\cite{das2022lilliput}, and Micro Blossom~\cite{wu2024micro}.
We next describe the network architecture of \name, which organizes the compute resources in a hybrid tree-grid topology for scalable and efficient decoding.

\subsection{Hybrid Tree-Grid Network Topology}

\begin{figure} [!t]
    \centering
        \includegraphics[width=0.99\linewidth]{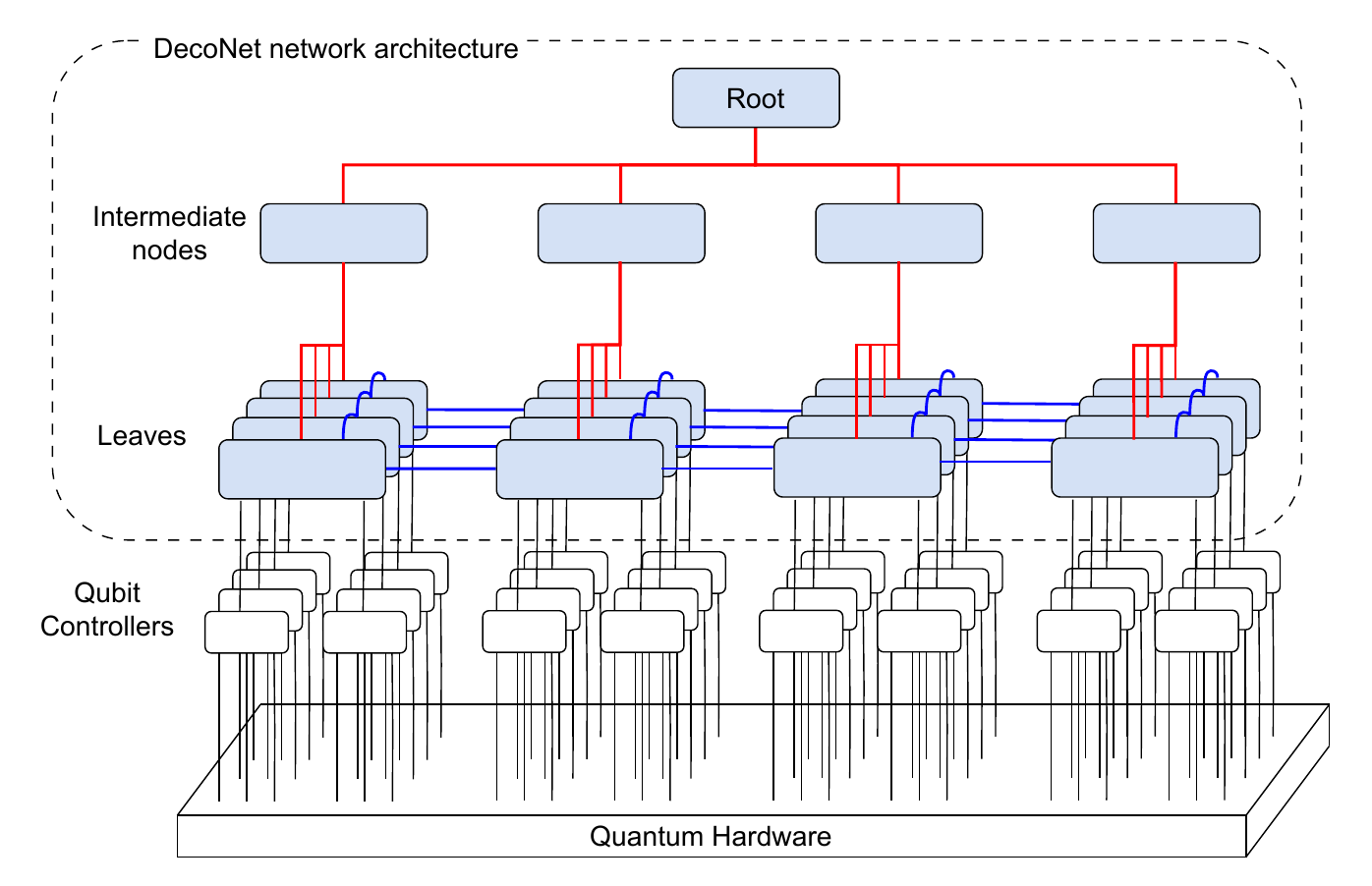}
        \caption{Network Architecture of \name, showing the hybrid tree-grid structure. The tree's leaf nodes run \units, while intermediary nodes route information.}
 	\label{fig:helios_tree}
\end{figure}

\autoref{fig:helios_tree} illustrates the hybrid tree-grid network topology employed by \name. 
Compute resources at the lowest level (leaf nodes) run \units, while intermediate nodes act as routers. 
The root node serves as the central interface for user interaction, configuration, and experiment monitoring.

This hybrid tree-grid topology brings the benefits of both tree and grid. 
The \emph{grid structure} efficiently handles local communication between adjacent leaf nodes. 
Through careful mapping of \blocks to leaf nodes (\S\ref{ss:mapping_to_leaf}), \name restricts boundary information exchange exclusively to adjacent compute resources in the grid.
By restricting boundary information to point-to-point grid connections, we eliminate routing overheads.

However, a purely grid-based structure becomes inefficient under worst-case communication scenarios, such as transferring logical states between distant nodes. 
Grid-only topologies result in a worst-case latency scaling of $O(\sqrt{n})$, where $n$ denotes the number of logical qubits. 
To mitigate this, we augment the grid with a \emph{tree structure}, reducing maximum communication latency between any two nodes to $O(\log(n))$. 
The tree structure handles both time-sensitive (logical results, control signals) and non-time-sensitive data (initial decoder configurations, monitoring signals).

We could not find a similar hybrid tree-grid topology in the literature, likely due to quantum computing having unique requirements: strong locality of information sharing for lattice surgery, where a grid topology excels, and the need to minimize worst-case latency, where a tree topology is ideal.

Furthermore, this topology supports scalable expansion of I/O resources. 
By directly connecting qubit controllers to leaf nodes, the system can scale its support for additional qubit controllers by incorporating more leaf nodes and, if necessary, adding intermediate routing nodes to expand the tree. Additionally, the same grid network can also facilitate the distribution of clock signals to qubit controllers.

To minimize worst-case latency, maintaining a minimal height for the \name tree structure is beneficial. The branching factor at each node and the necessary I/O capacity for qubit controllers, both dependent on the compute resource's maximum fan-out, dictate the tree's height. 
We leave the specific tree configuration as an implementation choice.

\subsection{Network Components of \name}

\subsubsection{Root Node}

The root node consists of three components: a user interface, a logical-level processor, and a router. 
The user interface facilitates system configuration, status monitoring, and end-user interaction, such as providing an input logical circuit. 
The logical-level processor executes the logical quantum circuit provided by the user and generates decoding instructions based on this circuit. 
The router distributes these instructions to leaf nodes and forwards messages from the leaf nodes to the logical-level processor.

\subsubsection{Leaf Nodes}

\begin{figure} [!t]
\centering
\includegraphics[width=0.65\linewidth]{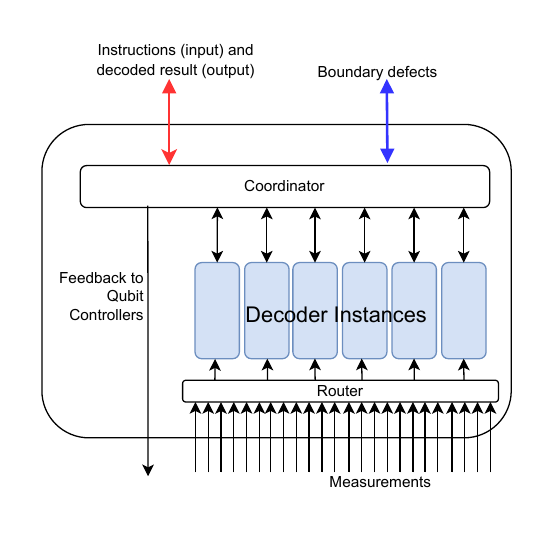}
\caption{Internal architecture of \name leaf nodes, comprising a coordinator, router, and multiple \units.}
\label{fig:leaf}
\end{figure}

\autoref{fig:leaf} shows the internal architecture of a leaf node, which comprises a coordinator, a router, and one or more \units. 
The coordinator controls the decoding workflow, interpreting and executing instructions from the root. 
This includes configuring boundaries between decoding blocks, initiating decoding operations on \units, and communicating boundary information between leaf nodes. 
We statically configure the router to route measurement data to the appropriate \units. 
The \units decode the blocks and combine them according to the boundaries between blocks.

\subsubsection{Intermediate Nodes}
Intermediate nodes route messages between the root node and leaf nodes.

\subsection{Mapping Decoder Instances to Compute Resources}
\label{ss:mapping_to_leaf}

We need to efficiently map decoding blocks to leaf nodes to minimize inter-resource communication. Our strategy of mapping is as follows. 

We statically assign each \unit to decode blocks corresponding to one logical qubit. 
This provides two benefits.
First, combining decoding blocks across the time domain does not require inter-resource communication.
Second, by connecting the relevant qubit controllers directly to this leaf node, we reduce measurement transmission latency, which contributes to the overall decoding latency.

We spatially group adjacent \blocks to the same node whenever possible, minimizing communication across the network. 
When we must spread adjacent decoding blocks across the network to multiple nodes due to resource constraints, we map them to compute resources that can directly communicate via the grid network, thereby localizing traffic and reducing overall network congestion.

\section{\hname Implementation}
\label{sec:implementation}

\begin{figure} [!t]
    \centering
        \includegraphics[width=0.90\linewidth]{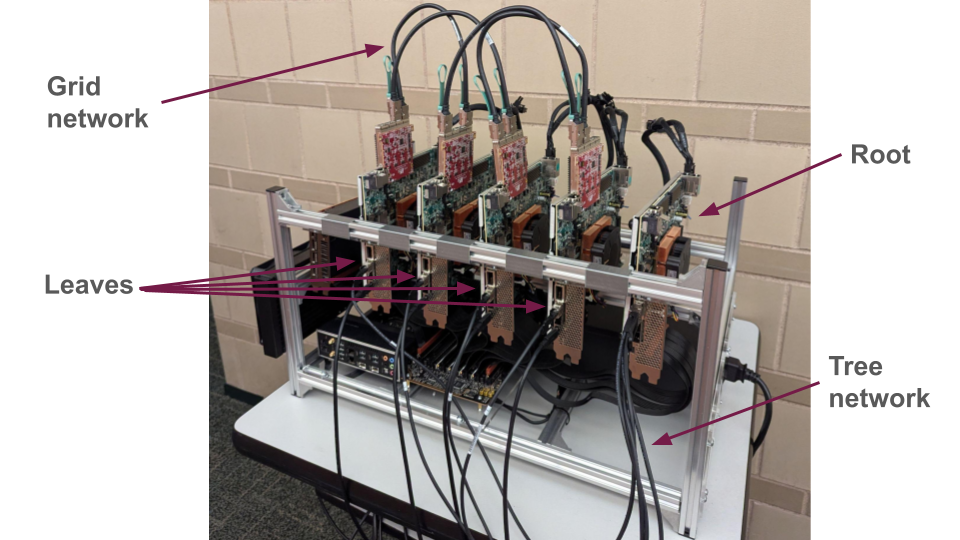}
        \caption{Implementation of \hname using five FPGAs. The rightmost FPGA is the root of the tree, and the other four FPGAs are leaf nodes.}
 	\label{fig:prototype}
\end{figure}

We built a concrete example of \name, called \hname, using five VMK180 evaluation boards, each featuring a Versal VM1802 FPGA-based SoC~\cite{vmk180}.
We chose these boards because they are one of the readily available Versal FPGA development boards with a very high LUT count and a high number of gigabit transceivers. 
\autoref{fig:prototype} shows the implementation \hname.
The network of our implementation is a two-level tree, with one root node and four leaf nodes, which we also connect in a grid.
We next describe the important choices made in our implementation. 

\subsection{Network Connection between FPGAs}

We implemented the FPGA interconnection using Gigabit Transceivers to maximize fan-out. 
Each VMK180 board includes 30 GTs exposed through QSFP, SFP+, and FMC+ links, allowing each parent node to connect to up to 25 child nodes. 
This high fan-out capability significantly reduces \name's tree height, enhancing scalability when the system grows.

The Aurora core, the standard Xilinx IP core designed for high-throughput communication~\cite{xilinxaurora}, incurs a high core-to-core latency of approximately 320 ns due to its optimization for throughput, making it unsuitable for real-time decoding. 
To address this limitation, we developed a custom low-latency transceiver, the Eos core~\cite{eoscoreGithub}, which reduces latency to 95 ns by sacrificing throughput. 
Eos core breaks messages into smaller chunks and inserts more frequent error correction codes, allowing the transmitter and receiver to process data in fewer cycles, minimizing overall latency. 
Despite the trade-off in throughput, the Eos core supports stable operation at 16 Gbps, which is sufficient for real-time decoding. We validated its reliability through 24-hour stress testing, confirming its suitability for \name. 
The Eos core is open-source and available at~\cite{eoscoreGithub}.

\subsubsection{Messaging format}

We use a fixed 64-bit format for messages exchanged between FPGAs. 
The first 8 bits specify the destination FPGA, the next 8 bits define the message header, and the remaining 48 bits serve as the payload. 
This simple format enables faster routing at each node, which is necessary for low-latency communication. 

\subsection{Choice of Decoder Instance: Helios}

We chose the Helios decoder as the \unit due to its favorable trade-off between scalability, latency, and accuracy~\cite{liyanage2024fpga}. Helios is a distributed implementation of the Union-Find (UF) decoder~\cite{delfosse2021almost}, which offers slightly lower accuracy than minimum-weight perfect matching (MWPM) decoders such as those used in Micro-Blossom~\cite{wu2024micro}. 
However, Helios achieves significantly lower latency and can decode surface codes up to $d=21$ in under 250~ns.
This low latency enables us to demonstrate that \name can support low-latency decoding. 

Since Helios is based on the Union-Find (UF) algorithm, we extended it to support the fusion operation, referred to as Fusion Union-Find (Fusion UF). We provide implementation details in \S\ref{sec:fusion}.

The choice of \unit presents a design trade-off. More accurate but resource-intensive decoders such as Micro-Blossom~\cite{wu2024micro} can improve logical error rates but increase decoding latency and reduce the number of logical qubits that can be supported per FPGA. 
In contrast, more resource-efficient UF-based decoders such as LCD~\cite{zaid2024local} can support more logical qubits per FPGA but incur higher decoding latency. Helios strikes a balance by offering low latency and moderate resource usage while maintaining acceptable decoding accuracy, making it suitable for our implementation.

\subsection{Implementation of FPGA logic}

We implemented the FPGA logic using Verilog and Tcl scripts, comprising approximately 9000 lines of code. 
The source code is publicly available at~\cite{deconetGithub}.
To support multi-qubit decoding, where boundaries dynamically change, we maintain an array of registers to track the state of each boundary. 
The coordinator has write access to these registers and updates them based on the instructions from the root. 
Each Helios instance within the FPGA has read access to these registers, allowing it to determine the current state of the boundary for the logical qubit it is decoding.

\subsection{Resource Usage}

\autoref{tab:FPGA_usage} presents the resource usage breakdown for the \hname implementation configured to decode 100 logical qubits at $d=5$. This configuration requires close to 1 million LUTs distributed across five FPGAs.

In the leaf nodes, \units consume approximately 95\% of the LUTs in the implementation, which is expected since decoding is the most compute-intensive task in the system. Adopting more resource-efficient \units, such as~\cite{barber2023realtime}, could potentially increase the system's decoding capacity per FPGA. We further explore these scalability limitations in \S\ref{ssec:eval_scalability}.

In contrast to the \units, the remaining logic in the leaf nodes consumes only around 14,000 LUTs and 21,000 registers, accounting for approximately 1.5\% of the FPGA’s resources. This small footprint leaves sufficient room for the compute-intensive \units. The 1.5\% includes the coordinator logic, inter-FPGA communication links, block-RAM-based FIFOs for inter-module communication, and peripheral logic for interacting with the ARM core. 

The root node utilizes only 2\% of the resources on the evaluation board, making it possible to use FPGAs with lower LUT counts for non-leaf nodes in \name. However, existing evaluation boards with smaller LUT capacities typically have fewer transceiver links, preventing us from pursuing this option. Using a board with fewer transceiver links would increase the tree height, leading to higher decoding latency. Alternatively, the unused resources in non-leaf nodes can potentially be repurposed for other tasks in the quantum control stack, such as logical qubit routing.

\begin{table}[!t]
    \centering
    \caption{Breakdown of resource usage for the configuration decoding maximum number of logical qubits, 100 ($d=5$)}
    \label{tab:FPGA_usage}
    \begin{tabular}{|l|r|r|r|}
        \hline
        \textbf{Component} & \textbf{LUTs} & \textbf{Registers} & \textbf{BRAMs} \\ \hline
        \multicolumn{4}{|c|}{\textbf{Root Node}} \\ \hline
        Eos core & 4389 & 5365 & 0 \\ \hline
        Root Coordinator & 53 & 646 & 0 \\ \hline
        Residual Logic & 14004 & 8274 & 46 \\ \hline
        \multicolumn{4}{|c|}{\textbf{Leaf Node (per FPGA)}} \\ \hline
        Coordinator & 25,038 & 3,859 & 0 \\ \hline
        Decoder Instances & 201,022 & 98,981 & 0 \\ \hline
        Eos core & 6141 & 7317 & 0 \\ \hline
        Residual Logic & 7753 & 13542 & 40 \\ \hline
    \end{tabular}
\end{table}

\subsection{Fusion Union-Find}
\label{sec:fusion}

We introduce Fusion Union-Find (Fusion UF), our novel approach for merging multiple partitions of a decoding graph, decoded using the Union-Find algorithm. 
This is an alternative to conventional merging techniques such as sliding window decoding and parallel window decoding~\cite{skoric2023parallel, tan2022scalable}.
Fusion UF draws inspiration from Fusion Blossom ~\cite{wu2023fusion}, which uses a similar methodology to speed up MWPM-based decoding. 

\subsubsection{Merging using Fusion Union-Find}

Fusion UF merges two blocks as follows. 
Initially, the Distributed Union-Find decoder processes the two blocks independently, treating their shared face as an artificial boundary. 
Any cluster with fully grown edges that touch this artificial boundary is considered even and do not grow further. 

After completing the clustering phase in both blocks,
the system removes this artificial boundary and calculates the cluster parities again. 
If any cluster in either block is odd, the decoder 
resumes the growing and merging phase in both blocks simultaneously until no more odd clusters remain. Finally, the system moves to peeling phase in each block. 

In \S\ref{ssec:fuf_latency}, we present empirical evidence demonstrating that Fusion UF achieves lower latency compared to other methods. 
\section{Evaluation}
\label{sec:evaluation}

The main objective of our evaluation is to assess the scalability of \name. To that end, we answer the following key questions:
\begin{itemize} [leftmargin=*]
    \item \textbf{Latency and throughput growth}: How do latency and throughput scale with the code distance and the number of merged logical qubits?
    \item \textbf{Limitations}: What are the scalability limits of \name?
    \item \textbf{Accuracy}: How does the fusion-UF approach compare in accuracy to alternative methods?
\end{itemize}

We first describe our methodology and then present empirical results answering these questions. To validate the system's practicality, we decode a set of micro benchmarks. 

\subsection{Methodology}

To evaluate the accuracy of fusion-UF, we extend the software simulation library~\cite{qecPlayground} to support fusion-UF. 
Software simulations enable us to test higher code distances ($d$) without encountering hardware resource constraints. 

For latency and throughput evaluations, we measure FPGA clock cycles required for syndrome decoding using \hname.
We define latency as the time between the availability of the last measurement round for a decoding block at the decoder and the availability of the decoded result. Inverse throughput is the time interval between a \unit accepting consecutive decoding blocks, normalized by $d$. 
We analyze the trends in latency and throughput to identify system limitations. 

\subsubsection{Experimental setup}

We use the five-FPGA Helios-Net implementation described in \S\ref{sec:implementation} as the experimental platform. The ARM cores on each evaluation board generate sample syndromes and transfer them to the \units in the FPGAs. To verify the correctness of our multi-FPGA implementation, we compare the logical state of each decoding block after every decoding round with results from offline simulations of the original UF decoder by Delfosse et al.~\cite{delfosse2021almost}, executed on the ARM core.
The comparison reveals identical results between the software simulation and the hardware implementation. We perform $10^6$ trials for each error rate and code distance. 

We use two configurations of the Helios decoder as \units in our evaluation. 
 In most experiments, we use the default Helios (Helios-1) configuration, which offers the best latency scalability with $d$. 
 To demonstrate the decoding of the maximum number of logical qubits, we use Helios-n, where $n=d$. 
 Helios-n requires fewer FPGA resources than Helios-1, supporting more decoding blocks on the implementation. 

We use $d=5$ as our default configuration, as we believe it would be a reasonable distance for the first experiments with multiple logical qubits on actual quantum hardware. 

\subsubsection{Noise Model}

We use the phenomenological noise model~\cite{dennis2002topological} with measurement errors for our experiments. 
Prior studies widely use this model for single-qubit decoders~\cite{liyanage2023scalable, das2022afs, holmes2020nisq+}. 
The system can be easily extended to other noise models, such as circuit-level noise and erasure errors, as the Helios decoder supports both models. 

For most of our experiments, we use a default noise level of $p=0.001$, consistent with prior  works~\cite{das2022afs, barber2023realtime, liyanage2023scalable}.  
This is a reasonable assumption as $p=0.001$ is more than 10 times below the threshold, which is necessary to exponentially reduce the logical error rates. 
Furthermore, for scalability evaluations, we randomly merge and split adjacent decoding blocks with a 50\% probability after every $d$ rounds. 

\subsubsection{Micro Benchmarks}

To validate the capabilities of our implementation, we create a set of microbenchmarks consisting of commonly required logical operations described in the literature that \name can decode. 

\subsection{Accuracy of Fusion-UF}
\label{ssec:fuf_accuracy}

\begin{figure}[!t]

\centering
 \subfloat[No notable accuracy loss]{
    \includegraphics[width=0.23\textwidth]{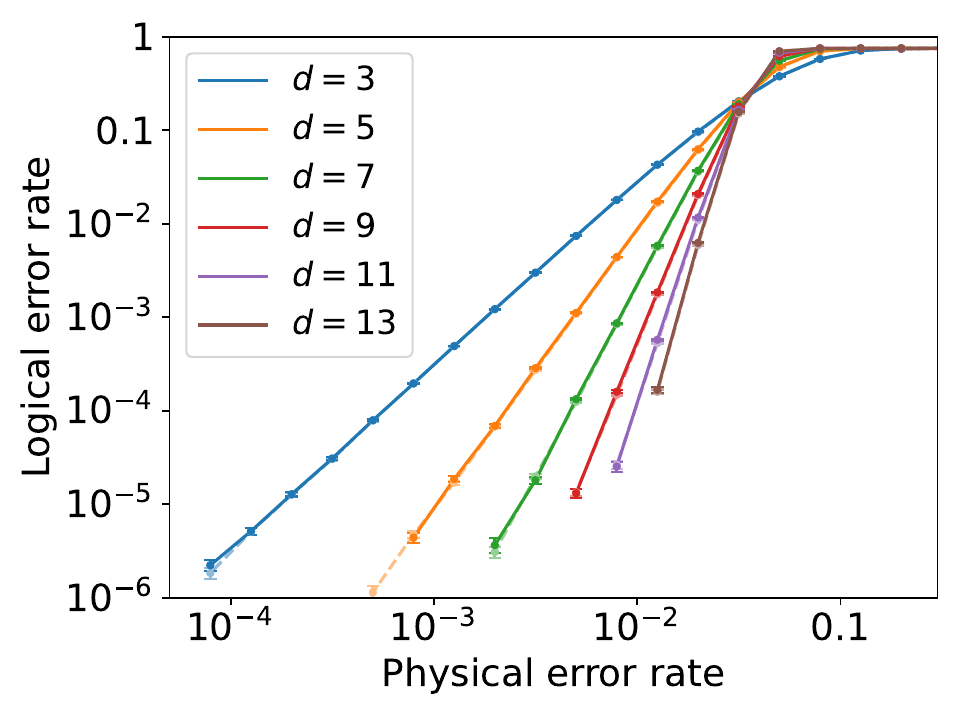}
    \label{fig:fusion_accuracy}
}
\centering
\subfloat[Lower latency]{
    \includegraphics[width=0.23\textwidth]{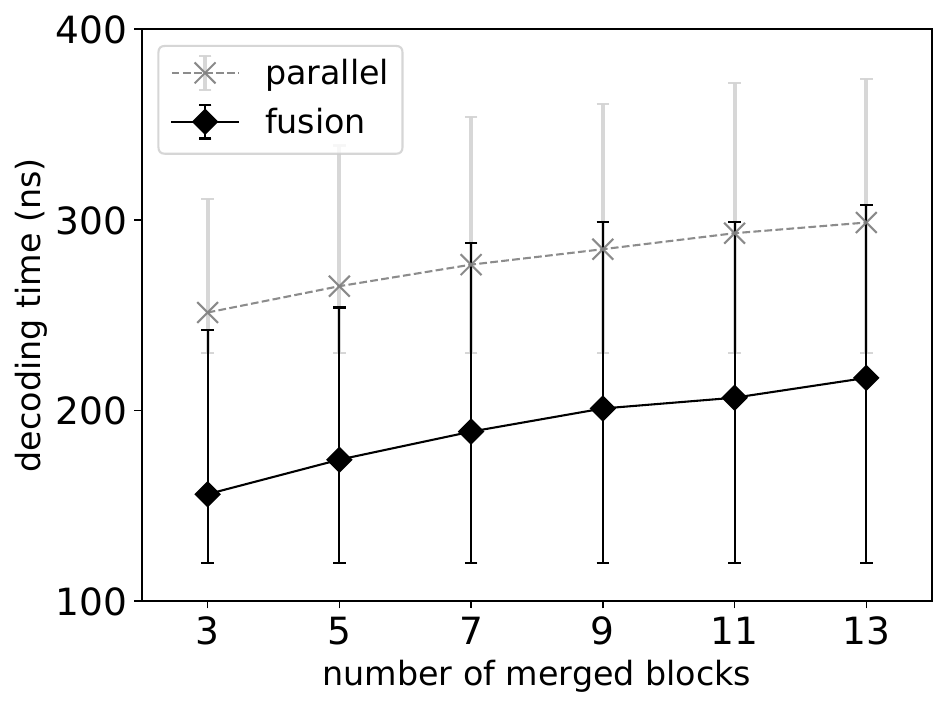}
    \label{fig:fusion_v_parallel}
}
\caption{Accuracy and scalability of the fusion-UF approach. (a) Comparison of decoding accuracy for two merged blocks using fusion-UF and a global UF decoder. Solid lines correspond to results from fusion-UF, and dashed lines represent results from a global decoder. Each data point aggregates $10^8$ trials. (b) Latency growth of fusion-UF compared to parallel window decoding for multiple merged logical qubits of $d=5$. The error bars indicate the variation of latency from the minimum to the $95^{th}$ percentile.}
\label{fig:fusion_eval}
  \end{figure}

To evaluate the accuracy of the fusion-UF approach, we decode two merged blocks with varying $d$ using fusion-UF and a global UF decoder, which considers both blocks as a single decoding graph. 
\autoref{fig:fusion_accuracy} illustrates the results. 
The results indicate that fusion-UF exhibits no statistically significant accuracy degradation, with a relative difference of less than 10\% compared to the global-UF decoder. 
Prior work on parallel window and sliding window decoding also reports no noticable accuracy loss~\cite{skoric2023parallel, lin2024spatially}.

\subsection{Latency of Fusion-UF} 
\label{ssec:fuf_latency}
We evaluate the scalability of Fusion-UF by comparing its decoding latency against parallel window decoding across varying numbers of merged logical qubits.
~\autoref{fig:fusion_v_parallel} shows the results, with the x-axis representing the number of merged logical qubits and the y-axis showing the decoding latency.
Fusion-UF achieves lower latency than parallel window decoding across all tested configurations. 
The reduction of latency is primarily due to increased parallelism and reduced redundant computation in Fusion-UF as mentioned in \S\ref{ssec:combining}.

\subsection{Scalability of \name}
\label{ssec:eval_scalability}

\begin{figure}[!t]
\edef\xfigwd{\the\columnwidth}

 \subfloat[Latency grows with $d$]{
    \includegraphics[width=0.47\linewidth]{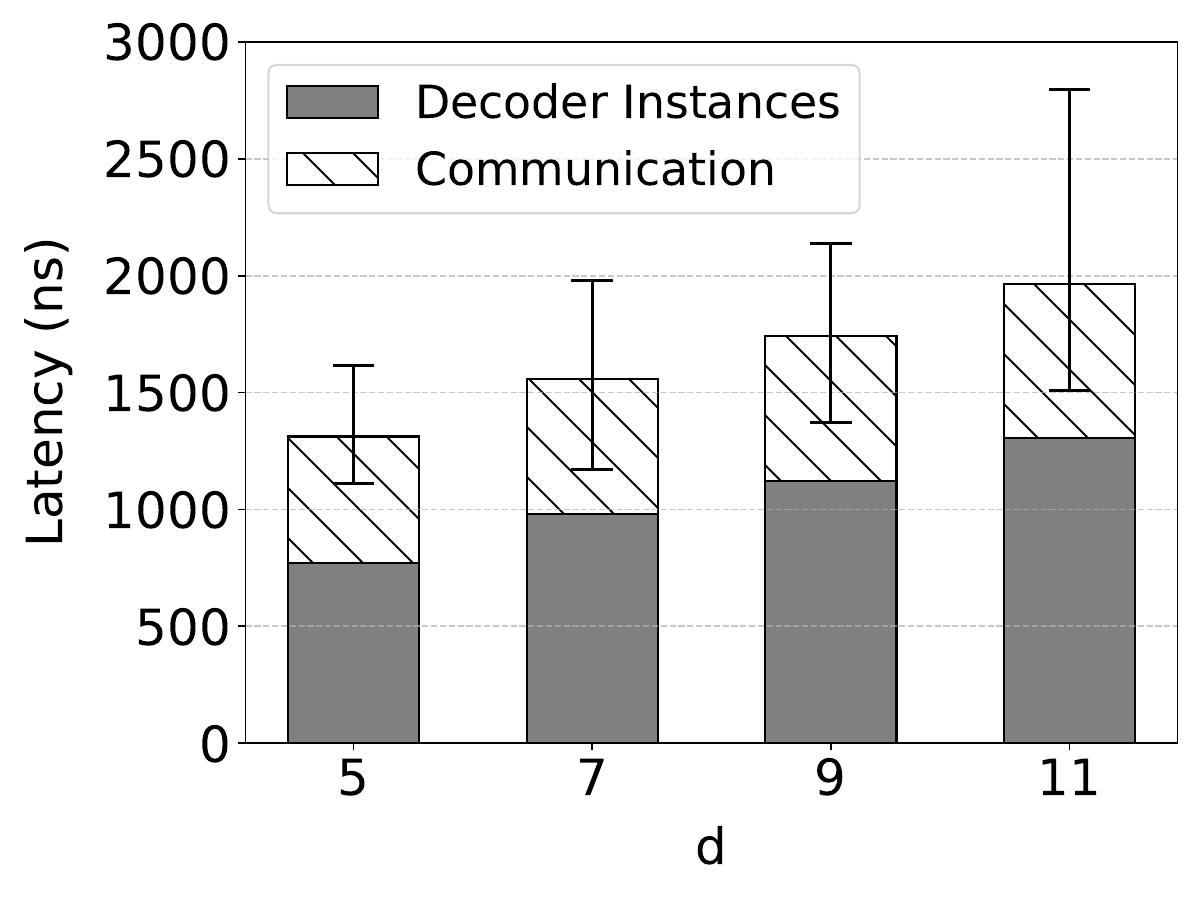}
    \label{fig:latency_v_d}
}
\subfloat[Latency grows with num qubits]{
    \includegraphics[width=0.47\linewidth]{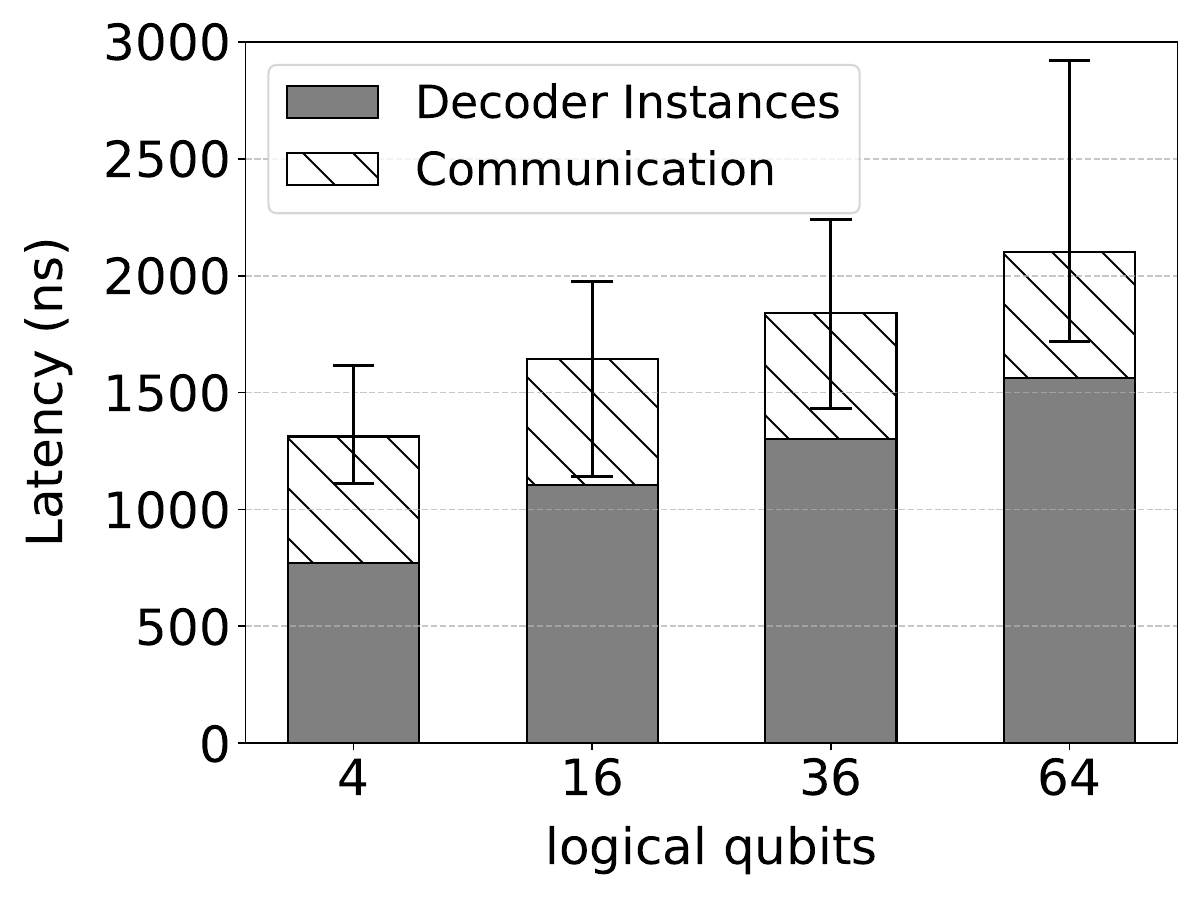}
    \label{fig:latency_v_lq}
}
\vspace{10pt}
\subfloat[I reduces with $d$]{
    \includegraphics[width=0.47\linewidth]{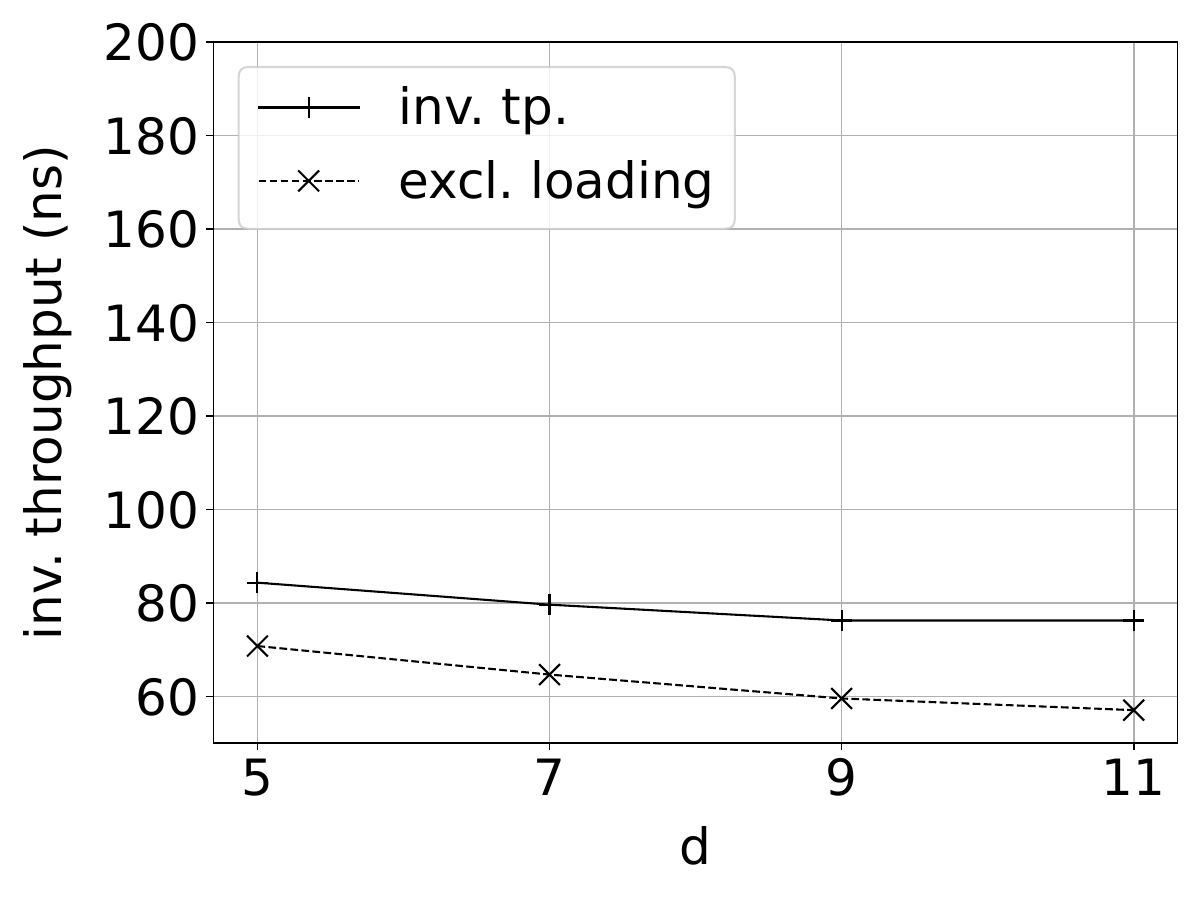}
    \label{fig:tp_v_d}
}
\subfloat[I grows with num qubits]{
    \includegraphics[width=0.47\linewidth]{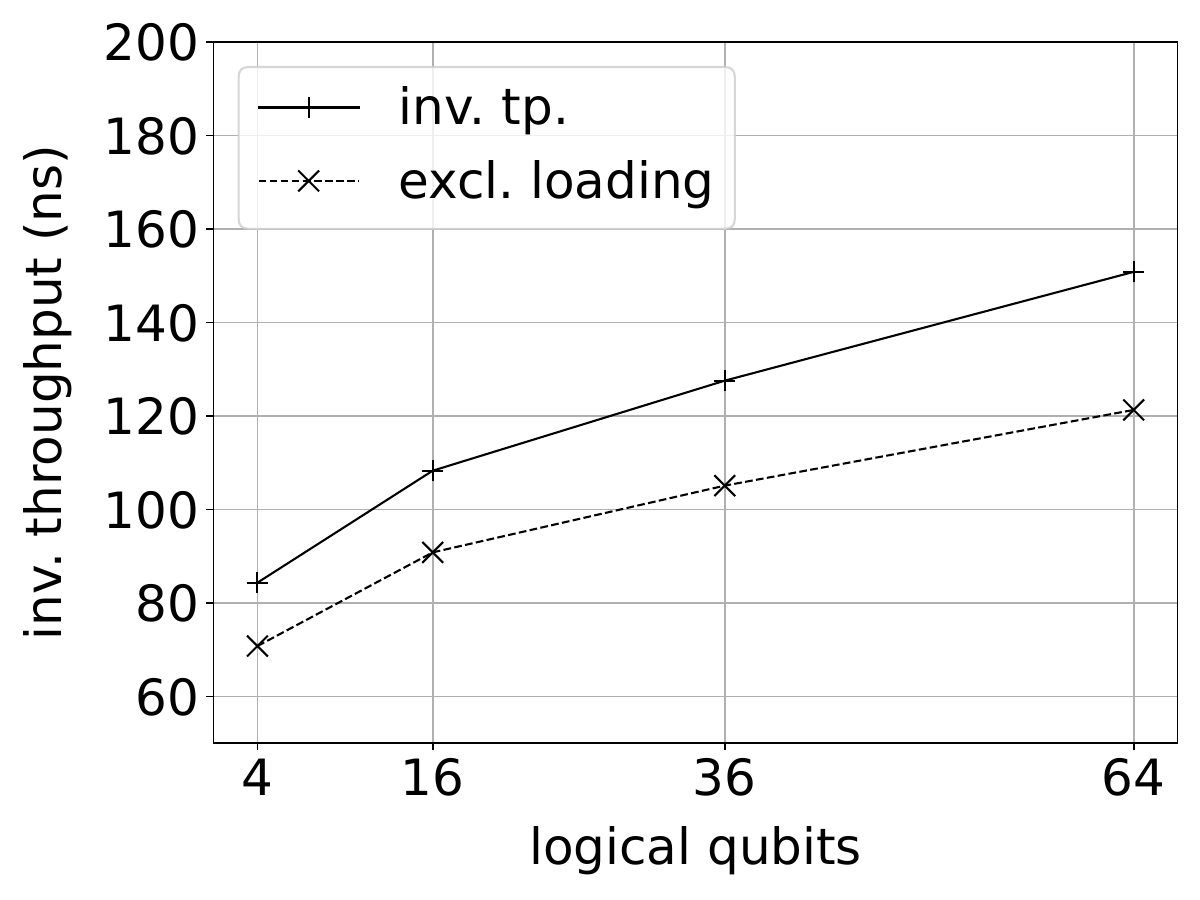}
    \label{fig:tp_v_l}
}
\caption{Scalability of \hname in terms of latency and inverse throughput (I). In (a) and (b), we plot the overall decoding latency as a function of code distance $d$ and the number of logical qubits per FPGA. We break down the total latency into contributions from \units and inter-FPGA communication. In (c) and (d), we show the inverse throughput (solid line) versus $d$ and the number of logical qubits. The error bars indicate the variation of total latency from the minimum to the $95^{th}$ percentile. We plot the inverse throughput, with and without defect data loading time (solid and dashed lines, respectively). (a) Latency increases with code distance. (b) Latency increases with the number of logical qubits per FPGA. (c) Inverse throughput remains at least $10\times$ faster than the measurement rate for all distances. (d) Inverse throughput increases with the number of logical qubits but remains $6\times$ faster than the measurement rate even when decoding 64 logical qubits. Default parameters: $d=5$, $p=0.001$, and 4 logical qubits.}
\label{fig:full_eval}
  \end{figure}

We evaluate the scalability of the \name using its implementation \hname with respect to $d$ and the number of logical qubits. We use two primary metrics: latency and inverse throughput.

\autoref{fig:latency_v_d} and \autoref{fig:tp_v_d} show how latency and inverse throughput scale with increasing $d$, when each leaf FPGA decodes one logical qubit. Across all supported code distances, the overall latency remains below the $d~\mu s$ measurement interval typical of superconducting quantum systems. As shown in \autoref{fig:latency_v_d}, latency consists of decoding within the \units and inter-FPGA communication. While decoder execution scales sublinearly with $d$, inter-FPGA communication scales as $O(d^2)$, since it depends on the boundary size between \blocks. However, this scaling trend is not prominent in our results due to a 400~ns lower-bound latency imposed by the cumulative delay across network links. 
\couldremove{
The variation in decoding latency is due to the variation in decoding time inside \units, due to the nature of distributed-UF decoding.
}

The inverse throughput improves slightly from 84 ns to 76 ns as $d$ increases from 5 to 11, consistent with trends observed in the Helios decoder for isolated qubits~\cite{liyanage2024fpga}. These values remain significantly below the 1~$\mu s$ measurement interval, confirming that \hname supports real-time decoding. Throughput is affected only by data loading and execution within a single \block, and remains independent of inter-FPGA communication latency due to the system’s ability to pipeline loading of the next decoding block while transmitting boundary information.

The current implementation loads defects sequentially from the ARM core. However, in practice, the defects would arrive from multiple qubit control devices, enabling parallel loading up to the fan-out of each leaf node. When we exclude data loading time from our measurements, the inverse throughput decreases from 71 ns to 57 ns as $d$ increases from 5 to 13.

\autoref{fig:latency_v_lq} and \autoref{fig:tp_v_l} show how latency and throughput scale with the number of logical qubits. Latency increases from 771 ns to 1562 ns as the number of logical qubits grows from 4 to 64. This latency remains under 20\% of the $d~\mu s$ measurement time. The inverse throughput increases from 71 ns to 121 ns, which remains well below the 1~$\mu s$ threshold.

Based on the experimental results showing that the inverse throughput scales sublinearly with $d$, the largest decoding block that can fit in an FPGA determines the maximum $d$. 
On VMK-180 SoCs, we support up to $d=13$. However, the design could potentially scale up to $d=23$ when using a VU19P FPGA, the largest commercially available option. 
Even though the number of decoding blocks per FPGA is also limited by logic utilization, this bottleneck is easily addressed by distributing blocks across additional leaf nodes.

The number of leaves and inter-FPGA latency do not impose immediate limits on scalability, as they impact latency but not throughput. However, increasing inter-FPGA latency reduces the logical operation frequency, leading to a polynomial increase in circuit execution time. 
Additionally, when the number of leaves increase, the system will eventually bottleneck at the root node, which determines circuit control paths based on prior decoding results. We can potentially avoid this bottleneck by taking distributed control decisions at intermediate nodes, but this direction requires further investigation.

\subsection{Decoding 100 logical qubits} 

We decode up to 100 logical qubits of $d=5$ using our five-FPGA implementation by employing the resource-efficient Helios-d configuration as the \unit. To support 100 logical qubits, each FPGA processes 25 logical qubits. Scaling beyond this point causes the decoding rate to fall below the measurement rate due to the limited scalability of the Helios-d configuration~\cite{liyanage2024fpga}.
When decoding 100 logical qubits, the system achieves an average latency of 12.01~$\mu$s and an inverse throughput of 0.84~$\mu$s. While these values are significantly higher than those of the Helios-1 configuration, they remain within the bounds required for real-time decoding.

In terms of scalability, the Helios-d configuration reaches the decoding time limit before exhausting FPGA resources. Based on latency growth trends, we estimate that for an error rate of $p=0.001$, Helios-d can support up to $d=25$.

\subsection{Microbenchmarks}

\begin{table}[!t]
    \centering
    \scriptsize
    \caption{Microbenchmark results showing latency and inverse throughput (standard deviation in brackets). Inverse throughput is normalized by $d$.}
    \label{tab:microbenchmarks}
    \begin{tabular}{|l|c|c|c|c|}
        \hline
        \textbf{Microbenchmark} & \textbf{\# L.} & \textbf{\#} & \textbf{Latency} & \textbf{Inv. Thpt}\\
        ~& \textbf{Qubits} & \textbf{Rounds} & \textbf{(ns)}& \textbf{(ns)}\\ \hline
        Meas.-based feedback & 1 & $d$ & 916 (191.1) & 86.9 (22.2) \\
        Merge + split & 2 & 3$d$ & 2003 (355.0) & 84.5 (35.1) \\
        Move qubit & 3 & 3$d$ & 2087 (238.9) & 87.2 (22.4) \\
        CNOT & 3 & 3$d$ & 3258 (619.9) & 86.5 (24.6) \\
        CNOT (plane layout) & 6 & 3$d$ & 3351 (484.8) & 91.5 (19.8) \\
        Single-ctrl multi-CNOT & 5 & 3$d$ & 3249 (482.7) & 91.7 (34.8) \\
        State expansion & 4 & 2$d$ & 2751 (536.4) & 86.4 (19.7) \\
        15-1 magic state distill. & 24 & 5$d$ & 5701 (633.0) & 123.1 (35.9) \\ \hline
    \end{tabular}
\end{table}

We report the decoding latency and inverse throughput of selected micro benchmarks in \autoref{tab:microbenchmarks}. Each entry lists the number of logical qubits and the number of measurement rounds required to implement the circuit and achieve fault-tolerant decoding. Because \name organizes decoding across FPGAs in a pipelined manner (\autoref{fig:decoding_pipeline}), some FPGAs process additional measurement rounds to complete the decoding of a logical circuit. 

All microbenchmarks exhibit an average decoding latency below 30\% of the measurement acquisition time. The inverse throughput is 8$\times$ faster than the measurement rate of 1~$\mu s$, ensuring that \hname operates backlog-free across all benchmarks. We highlight two microbenchmarks that demonstrate key capabilities of \name: measurement-based feedback and 15-1 magic state distillation.

In measurement-based feedback, the system decodes a logical qubit and transmits the result to another FPGA, a control flow required in the T-gate.
The quantum hardware must idle until the result is available, and circuit execution can slow polynomially due to this latency~\cite{skoric2023parallel}. In our experiments, the average latency for decoding and transmitting the result to another FPGA is 0.91~$\mu$s, more than five times faster than the time to measure $d$ rounds. Across $10^6$ trials, we observe a worst-case latency of 2.25~$\mu$s, which is still over twice as fast as the time to measure $d$ rounds. As \hname can consistently deliver results before the next measurement round becomes available, it introduces minimal slowdown to quantum circuit execution.

15-1 magic state distillation, spanning 24 logical qubits, represents one of the largest merge-and-split-based operations required for FTQC. This circuit is essential for generating T-gates and involves five rounds of CNOT gates~\cite{fowler2019lowoverhead}. A practical decoder must process this circuit faster than the rate of measurement to be useful for large-scale FTQC. \name decodes the corresponding 5$d$ rounds in 5.7~$\mu$s for $d=5$, achieving inverse throughput $8\times$ faster than the rate of measurement. 

The decoding latency of each microbenchmark also depends on how many FPGAs participate in decoding, which is determined by the placement of the logical circuit. To evaluate worst-case behavior, we map each circuit across FPGA boundaries whenever possible, maximizing inter-FPGA communication. Mapping a circuit to a single FPGA reduces latency. For example, a CNOT gate requires 1297 ns on a single FPGA versus 3258 ns across three. However, inverse throughput remains statistically unchanged due to \name’s pipelined execution across all FPGAs.

\subsection{Comparison with related work}

We compare \hname with QULATIS~\cite{ueno2022qulatis}, the only prior hardware decoder design in the literature that provides detailed decoding latencies. 

\hname achieves significantly higher decoding accuracy than QULATIS, while QULATIS outperforms \hname in latency and throughput. 
At $p=0.001$ and $d=5$, \hname provides over two orders of magnitude better accuracy than QULATIS, and this gap widens with increasing $d$. 
This is due to QULATIS using a greedy decoding algorithm, which has orders of magnitude lower accuracy than Union-Find.
QULATIS achieves lower latency due to its higher operating frequency. For 15-1 magic state distillation at $d=5$ and $p=0.001$, QULATIS reports a latency of 1.16~$\mu$s based on SPICE-level simulation at 2 GHz, whereas \hname achieves 5.7 $\mu$s when running at 100 MHz. In terms of cycle count, QULATIS requires 2235 cycles, while \hname completes decoding in 570 cycles. For the same circuit, QULATIS achieves an inverse throughput of 46.7 ns, compared to 123.1~ns for \hname, again primarily due to the clock frequency difference.

\balance
QULATIS also faces more stringent scalability limits. Its power consumption at cryogenic temperatures (4K) limits the number of supported logical qubits. Their estimation suggests it can run 40 concurrent 15-1 distillation circuits of $d=9$. In contrast, \name’s scalability is limited by the root node’s capacity to process decoded results, a bottleneck that does not emerge until scaling to thousands of logical qubits.

\section{Conclusion}

We present \name, a network-integrated decoding system for fault-tolerant quantum computers that decodes a dynamic graph of multiple interacting logical qubits. 
\name introduces a scalable architecture that expands compute and I/O resources by trivially adding hardware, enabling the decoding of thousands of logical qubits. 
Using a five-FPGA implementation, \hname decodes 100 logical qubits of distance five in real-time.
To the best of our knowledge, this is the highest number of interacting logical qubits decoded faster than the measurement rate by any system.
Given this scalability, we consider \name a strong candidate for the logical qubit decoding layer in future fault-tolerant quantum computers.

\section*{Acknowledgment}
This work was supported in part by Yale University and NSF MRI Award \#2216030.


\clearpage
\bibliographystyle{IEEEtran}
\balance
\bibliography{abr-long, qec, ref2}{}


\end{document}